\documentclass[aps,pra,superscriptaddress,amsmath,amssymb,preprintnumbers]{revtex4}
\usepackage{amssymb}
\usepackage{booktabs}
\usepackage{graphicx}
\usepackage{bm}
\usepackage{color}
\usepackage{cancel}
\usepackage{slashed}

\usepackage{subfigure}

\newcommand{\Ignore}[1]{}
\newcommand{\Delete}[1]{{\color{red}[#1]}}
\renewcommand{\Delete}[1]{}

\newcommand{\Ket}[1]{\left\vert #1\right\rangle}
\newcommand{\Bra}[1]{\left\langle #1\right\vert}
\newcommand{\PKet}[1]{\vert #1\rangle}
\newcommand{\PBra}[1]{\langle #1\vert}
\newcommand{\BraKet}[2]{\left\langle#1\vert #2\right\rangle}
\newcommand{\KetBra}[2]{\left\vert#1\right\rangle\left\langle#2\right\vert}
\newcommand{\PBraKet}[2]{\langle #1 \vert #2 \rangle}

\newcommand{\ii}{\mathrm{i}}
\newcommand{\ee}{\mathrm{e}}

\newcommand{\bathsize}{{L}}
\newcommand{\auxt}{w}
\newcommand{\auxtt}{w'}

\newcommand{\IdOp}{1\!\!1}
\newcommand{\opidentity}{\IdOp}
\newcommand{\alldown}{\{ \downarrow \}}
\newcommand{\alldownbutone}[1]{ \{ \downarrow \} \uparrow_{#1} \{ \downarrow \} }
\newcommand{\alldownbuttwo}[2]{ \{ \downarrow \} \uparrow_{#1} \{ \downarrow \} \uparrow_{#2} \{ \downarrow \} }

\newcommand{\bitilde}[1]{\overline{#1}}
\newcommand{\tritilde}[1]{\mathring{#1}}

\newcommand{\rwa}{{(0)}}
\newcommand{\crt}{{(\omega)}}

\newcommand{\hamfintld}{\bitilde{H}_G}

\newcommand{\CorrOne}{\Lambda}

\begin{document}


\pagecolor{white}


\title{Adiabatically-manipulated systems interacting with spin baths beyond the Rotating Wave Approximation}


\author{Benedetto Militello}
\author{Anna Napoli}
\affiliation{Universit\`a degli Studi di Palermo, Dipartimento di Fisica e Chimica - Emilio Segr\`e, Via Archirafi 36, 90123 Palermo, Italia}
\affiliation{I.N.F.N. Sezione di Catania, Via Santa Sofia 64, I-95123 Catania, Italia}


\begin{abstract}
The Stimulated Raman Adiabatic Passage on a three-state system interacting with a spin bath is considered focusing on the efficiency of the population transfer. Our analysis is based on the perturbation treatment of the interaction term evaluated beyond the Rotating Wave Approximation, thus focusing on the limit of weak system-bath coupling. The analytical expression of the correction to the efficiency and consequent numerical analysis show that in most of the cases the effects of the environment are negligible, confirming the robustness of the population transfer.
\end{abstract}

\maketitle 


\section{Introduction}\label{sec:introduction}

A quantum system ruled by a slowly varying Hamiltonian undergoes a dynamics known as adiabatic following of the eigenstates, based on the adiabatic theorem, according to which the populations of the instantaneous eigenstates of the Hamiltonian do not change~\cite{ref:Messiah,ref:Griffiths} (details are recalled in \ref{app:Adiabatic}). This physical behavior is the key ingredient of many protocols aimed at controlling a quantum system~\cite{ref:Lidar2018,ref:Santos2020,ref:Ilin2021,ref:Pys2022,ref:Coello2022}. 
Stimulated Raman Adiabatic Passage (STIRAP) ~\cite{ref:STIRAP_rev1,ref:STIRAP_rev2,ref:STIRAP_rev3,ref:STIRAP_rev4,ref:STIRAP_rev5} represents an important and well known example of adiabatic process. 
The STIRAP technique was introduced to realize a complete population transfer from one state to another one by exploiting a Raman scheme involving suitable pulses with time-dependent amplitudes which couple each of the two previously mentioned states with an auxiliary one.  The pulses have to be set in such a way that they should ensure the validity of the adiabatic approximation. Moreover, their specific time-dependence must be such that an eigenstate of the Hamiltonian coincides with the initial state of the system at the beginning of the process and with the target state at the end of the application of the pulses. Therefore, differently from what one could think at first glance, the total population transfer from the initial state to the target state is not due to radiative two-photon processes but plainly by an adiabatic following of an Hamiltonian eigenstate. This is particularly evident when the so called counter-intuitive sequence is considered, where the pulse which couples the auxiliary state and the target one precedes the pulse which couples the initial state to the auxiliary one. Indeed, in such a situation it is not reasonable to interpret the whole process as a (possibly virtual) photon absorption concomitant to the initial-auxiliary state transition followed by a (possibly virtual) photon emission concomitant to the auxiliary-target state transition. Moreover, it is worth observing that not only the counter-intuitive sequence works well, but that it usually works better than the so called intuitive sequence where the coupling between the auxiliary state and the initial one precedes the other coupling. (A detailed analysis of the counterintuitive sequence is given in sec.~\ref{sec:IdealStirap}.)

The STIRAP technique is still extensively investigated~\cite{ref:Blekos2020,ref:Ahmadinouri2020,ref:Cinins2022,ref:Dogra2022,ref:Liu2023,ref:Genov2023} and has been exploited in different physical contexts ranging from cold gases~\cite{ref:Ni2008,ref:Danzl2008,ref:Danzl2010} to condensed matter~\cite{ref:Halfmann2007,ref:Alexander2008,ref:Golter2014,ref:Yale2016,ref:Baksic2016,ref:Zhou2017,ref:Wolfowicz2021}, plasmonic systems~\cite{ref:Varguet2016,ref:Castellini2018}, superconducting devices~\cite{ref:Kubo2016,ref:Xu2016,ref:Kumar2016}, trapped ions~\cite{ref:Soresen2006,ref:Higgins2017} and optomechanical systems~\cite{ref:Fedoseev2021}. Recently, in order to improve the original technique by shortening the population transfer process, modifications of the original scheme including shortcuts to adiabaticity have been proposed~\cite{ref:Guery2019,ref:Vitanov2020,ref:Stefanatos2020,ref:Stefanatos2022,ref:Messikh2022,ref:Evange2023}. However, in this case a more complicated apparatus is required, which constitutes a disadvantage with respect to the original scheme. Indeed, implementation of shortcuts requires activation of additional interactions which somehow compensate any deviation from the adiabatic following of the eigenstates of the Hamiltonian, even in the cases where the Hamiltonian does not change slowly. In this way, the original time-dependent Hamiltonian of the system, say $H(t)$, does not rule anymore the dynamics, since also new terms, say $\delta H(t)$, are to be considered, and the total Hamiltonian $H(t) + \delta H(t)$ induces a unitary evolution which coincides with an adiabatic following of the eigenstates of $H(t)$ only. The structure of $\delta H(t)$ is generally complicated and the relevant pulses need to be very precise.

As a general fact, a quantum system is subjected to the effects of noise either rising from the interaction with external systems, as for example the constituents of the environment, or related to imperfections of the apparatus. Therefore, on the one hand the fidelity of the population transfer with respect to uncertainty or fluctuations of amplitudes and phases of the pulses has been analyzed~\cite{ref:Genov2013,ref:Yatsenko2014}. On the other hand, the interaction with the quantized electromagnetic field has been considered, for example exploiting effective non-Hermitian Hamiltonians, which is limited to the case where the states involved in the STIRAP scheme decay toward states not involved in the procedure~\cite{ref:Vitanov1997}. Beyond such a specific scenario, a more general approach based on the theory of open quantum systems is required~\cite{ref:Petruccione,ref:Gardiner}. 
In this line, master equations have been proposed to study the STIRAP-manipulated systems~\cite{ref:Ivaniv2005,ref:Mathisen2018,ref:New1,ref:New2}.  Moreover, under suitable assumptions fitting the Davies and Spohn theory for open quantum systems ruled by time-dependent Hamiltonians~\cite{ref:DaviesSpohn}, time-dependent master equations in the Lindblad form can be obtained from a microscopic interaction model between the atomic system and the quantized field~\cite{ref:Scala2010,ref:Scala2011}.

Since in many cases the system to be manipulated is close to other atomic systems and interacting with them, it can happen that the interaction with the electromagnetic field is not the main source of quantum noise or, at least, not the only significant. Nitrogen Vacancies in diamond~\cite{{ref:Golter2014,ref:Baksic2016,ref:Zhou2017}} or rare-earth doped crystals~\cite{ref:Halfmann2007,ref:Alexander2008} are two typical examples of such a scenario. 
Manipulation of spin defects in magnetic materials through adiabatic following-based techniques has been recently studied in the presence of interaction with the surrounding spins~\cite{ref:Onizhuk2021}. Moreover, very recently,  STIRAP processes on a system interacting with a spin bath has been theoretically analyzed under the assumptions that allow for the Rotating Wave Approximation (RWA) in the interaction between the three-state system and the spins of the environment~\cite{ref:MiliNapo2023}.  This study has been developed through evaluation of the unitary dynamics, in spite of the fact that master equations can be derived also for spin environments~\cite{ref:Fischer2007,ref:Ferraro2008,ref:Bhatta2017}.

In this paper, we extend the previous study in Ref.~\cite{ref:MiliNapo2023} still exploiting unitary evolution of the universe but overcoming the RWA in the system-environment interaction, which makes the physical model more realistic. In fact, while the RWA implies conservation of the total number of excitation (a feature which has been extensively used in the previous analysis), it somehow excludes a variety of possible transitions. On the contrary,  in this work we take into account all the terms of the system-bath interaction, thus introducing processes which can be responsible for a reduction of the population transfer efficiency.
The paper is structured as follows: in the next section we describe the physical system and the relevant Hamiltonian model, also providing a brief sketch of the STIRAP technique in the ideal case and beyond, also introducing the theoretical analysis based on the perturbation approach, specifically to the truncation of the Dyson series, after two changes of picture. In sec.~\ref{sec:results} we give the explicit form of the correction to the efficiency of the population transfer according to our theory and then we show predictions based on numerical calculations. Finally, in sec.~\ref{sec:discussion} we present an extensive discussion on the results. Two appendixes complete the presentation: in the first one the adiabatic approximation is recalled while in the second appendix we provide details about the matrix elements involved in the perturbation treatment.


\section{Physical Model and Methods}\label{sec:PhysicalModel}

\subsection{Hamiltonian Model}\label{sec:ham_mod} 

\begin{figure}[h!]
\includegraphics[width=0.6\textwidth, angle=0]{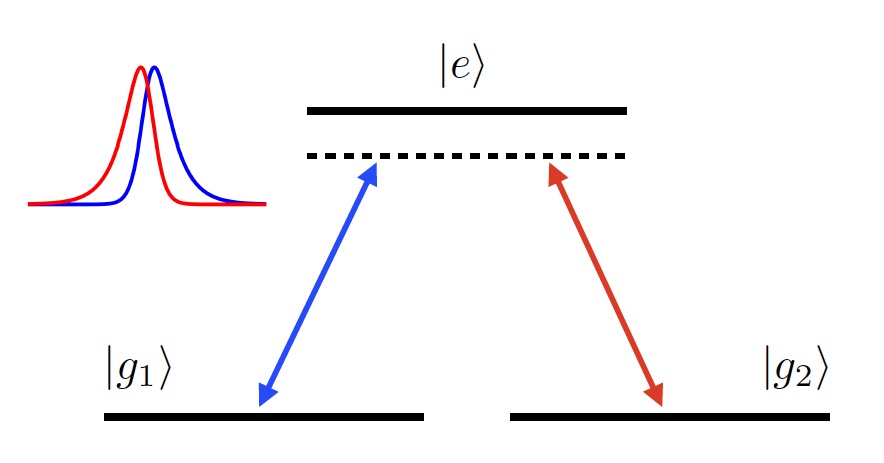}
\caption{STIRAP scheme: two lower states ($\Ket{g_1}$ and $\Ket{g_2}$) are coupled to an upper one ($\Ket{e}$) through suitable pulses. The inset shows the typical shape of the pulses.} 
\label{fig:STIRAP_scheme}
\end{figure}

The physical system we are focusing on consists of a three-state system subjected to two coherent fields and interacting with the surrounding environment consisting of a spin bath. The three-state system has two ground states ($\Ket{g_1}$ and $\Ket{g_2}$) and an excited state ($\Ket{e}$), and its free dynamics is governed by the Hamiltonian $H_A$ given below. The action of the STIRAP pulses coupling each of the two ground states with the excited one (see Fig.~\ref{fig:STIRAP_scheme}) is described by $H_S$. In addition, the free spin-bath (an ensemble of two-state systems) is described by $H_B$ while the system-bath interaction is described by $H_{AB}$ which associates spin flips with atomic transitions between the excited state and each of the two ground states. Therefore, the total Hamiltonian is given by ($\hbar=1$):
\begin{subequations}
\begin{equation}\label{eq:HTot}
H = H_A + H_S + H_B + H_{AB}\,,
\end{equation}
with:
\begin{eqnarray}
H_A &=& \nu \Ket{e}\Bra{e} 
\end{eqnarray}
\begin{eqnarray}
H_S &=& \sum_{m=1}^2 \Omega_m(t) \, \cos(\nu't) \, (\Ket{g_m}\Bra{e} + \Ket{e}\Bra{g_m}) \,,
\end{eqnarray}
\begin{equation}
H_B = \sum_{k=1}^\bathsize \frac{\omega}{2} \sigma_z^{(k)} \,,
\end{equation}
\begin{eqnarray}\label{eq:HabDef}
\nonumber
H_{AB} &=& \sum_{m=1}^2 \left(\Ket{g_m}\Bra{e}+\Ket{e}\Bra{g_m}\right)\otimes\sum_{k=1}^\bathsize\eta_k^{(m)} \sigma_x^{(k)} \,,\\
\end{eqnarray}
\end{subequations}
where $\nu$ is the energy gap between the free excited states and the two grounds, $\nu'$ is the frequency of the two pulses whereas $\Omega_m$'s describe the profiles of the pulses; the natural frequency of the spins of the bath is $\omega$, while the quantity $\eta_k^{(m)}$ is the spin-atom coupling constant related to the $k$-th spin and to the atomic transitions $\Ket{e} \leftrightarrow \Ket{g_m}$. Finally $\sigma_\alpha^{(k)}$ are the Pauli operators associated to the $k$-th spin.

Introducing the Hermitian operator $G_1=\nu' \Ket{e}\Bra{e} +H_B$ and the relevant unitary operator $U_1$ such that $\ii \dot{U}_1 = G_1 U_1$, in the new picture given by $\PKet{\tilde{\psi}(t)} = U_1^\dag(t) \Ket{\psi(t)}$ and $\tilde{A}(t) = U_1^\dag(t) A(t) U_1(t)$, we have that the generator of the time evolution is:
\begin{subequations}
\begin{equation}\label{eq:HTot_IP}
\tilde{H}(t) - G_1 = H_A' + \tilde{H}_S^\rwa(t) + \tilde{H}_S^\crt(t) + \tilde{H}_{AB}(t)  \,,
\end{equation}
with 
\begin{eqnarray}
H_A' = (\nu - \nu') \Ket{e}\Bra{e}  \,,
\end{eqnarray}
\begin{eqnarray}
\tilde{H}_S^\rwa(t) &=& \sum_{m=1}^2 \Omega_m(t) (\Ket{g_m}\Bra{e} + \Ket{e}\Bra{g_m}) \,,
\end{eqnarray}
\begin{eqnarray}
\tilde{H}_S^\crt (t) &=& \sum_{m=1}^2 \Omega_m(t) \Ket{g_m}\Bra{e} \ee^{- \ii 2 \nu' t}  + h.c.\,, \,\,\,\,\,\,\,\,\,\,
\end{eqnarray}
\begin{eqnarray}
\tilde{H}_{AB}(t) &=&\sum_{m=1}^2 (\Ket{g_m}\Bra{e} \ee^{- \ii \nu' t}  + \Ket{e}\Bra{g_m} \ee^{ \ii \nu' t} ) \otimes 
\sum_k \eta_k^{(m)} (\sigma_-^{(k)} \ee^{- \ii \omega t} + \sigma_+^{(k)} \ee^{ \ii \omega t})  \,, \,\,\,
\end{eqnarray}
\end{subequations}
with  $\Delta=\nu-\nu'$ the detuning between the atomic frequency and the field frequency. It is worth mentioning that, in view of the further treatment, we have split the transformed pulse Hamiltonian $\tilde{H}_S$ in two contributions, $\tilde{H}_S=\tilde{H}_S^\rwa+\tilde{H}_S^\crt$, where the first corresponds to the so called rotating terms (characterized by the absence of fast oscillations) while the second is related to the counter-rotating terms (rapidly oscillating). For our treatment of the system-bath interaction terms in $\tilde{H}_{AB}(t)$ this separation of rotating and counter-rotating terms is not necessary.

\subsection{Ideal STIRAP}\label{sec:IdealStirap}

Let us first sum up the basic of the ideal STIRAP, which corresponds to the absence of interaction with the environment (referring to our model, this condition is accomplished assuming $\eta_k^{(m)}=0$\,, $\forall m, k$). Moreover, assuming high atomic and field frequencies, one is legitimated to neglect the counter-rotating terms in the STIRAP Hamiltonian, so that the system can be assumed to be approximately ruled only by $H_A'+\tilde{H}_S^\rwa$, in this new picture usually addressed as the rotating frame. 
The operator
\begin{eqnarray}
H_A' + \tilde{H}_S^\rwa(t) = \Delta \Ket{e}\Bra{e}  + \sum_{m=1}^2 \Omega_m(t) (\Ket{g_m}\Bra{e} + \Ket{e}\Bra{g_m})\,,
\end{eqnarray}
can be easily diagonalized at every instant of time, and its instantaneous eigenstates are:
\begin{subequations}
\begin{eqnarray}
\Ket{+(t)} &=& \sin\varphi(t)\sin\theta(t)\Ket{g_1}+\cos\varphi(t)\Ket{e}+\sin\varphi(t)\cos\theta(t)\Ket{g_2}  \\
\Ket{0(t)} &=& \cos\theta(t)\Ket{g_1}-\sin\theta(t)\Ket{g_2}    \\
\Ket{-(t)} &=& \cos\varphi(t)\sin\theta(t)\Ket{g_1}-\sin\varphi(t)\Ket{e}+\cos\varphi(t)\cos\theta(t)\Ket{g_2}  
\end{eqnarray}
\end{subequations}
where
\begin{subequations}
\begin{align}
\label{eq:thetadef}
\tan\theta(t)&=\frac{\Omega_1(t)}{\Omega_2(t)}\,,\\
\label{eq:phidef}
\tan{2\varphi(t)}&=\frac{2\Omega_L(t)}{\Delta}\,,\\
\Omega_L(t)&=\sqrt{\Omega_1(t)^2+\Omega_2(t)^2}\,.
\end{align}
\end{subequations}

Such eigenstates correspond to the following eigenvalues:
\begin{eqnarray}
E_0 = 0 \,, \qquad E_{\pm} = \frac{1}{2} \left[\Delta \pm \sqrt{\Delta^2 + 4 \Omega_L(t)^2} \, \right]\,.
\end{eqnarray}

The standard STIRAP process aimed at transferring population from the state $\Ket{g_1}$ to $\Ket{g_2}$ is realized through a so called counterintuitive sequence, where the pulse $\Omega_2$ precedes $\Omega_1$. Accordingly, at the initial time one has $\theta = 0$ and $\varphi=0$, so that $\Ket{+} = \Ket{e}$, $\Ket{0} = \Ket{g_1}$ and $\Ket{-} = \Ket{g_2}$, while in the final time one has $\theta=\pi/2$ and $\varphi=0$, so that $\Ket{+} = \Ket{e}$, $\Ket{0} = -\Ket{g_2}$ and $\Ket{-} = \Ket{g_1}$. When the pulse profiles are slowly varying functions, the hypotheses of the adiabatic theorem are satisfied and the population of each eigenstate is preserved during the evolution. Consequently, in particular, the population of the state $\Ket{g_1}$, initially coinciding with $\Ket{0}$, is totally transferred to the state $\Ket{g_2}$, which equals $\Ket{0}$ in the final time.

\subsection{Time Evolution and Efficiency}\label{sec:time_evol}

In order to better evaluate the effects of the remaining terms of the Hamiltonian, we perform a new change of picture, applying the transformation $\Ket{\bitilde{\psi}} = U_2^\dag \Ket{\tilde{\psi}}$ with $U_2$ such that $\ii \dot{U}_2 = ( H_A' + \tilde{H}_S^\rwa\!(t) ) U_2$. The operator $U_2$ is well approximated by the unitary evolution describing the adiabatic following of the instantaneous eigenstates of the Hamiltonian $H_A' + \tilde{H}_S^\rwa\!(t)$, given by $U_2 (t, t_0) \approx \sum_m \ee^{-\ii \alpha_m(t, t_0)} \Ket{\phi_m(t)}\Bra{\phi_m(t_0)}$, with $\Ket{\phi_m(t)}$ the instantaneous eigenstates of $H_A' + \tilde{H}_S^\rwa(t)$ and $\alpha_m(t,t_0) = \int_{t_0}^t E_m(\auxt) \mathrm{d}\auxt + \ii \int_{t_0}^t (\partial_\auxt \Bra{\phi_m(\auxt)}) \Ket{\phi_m(\auxt)}\mathrm{d}\auxt $ the relevant phase factors. It is worth mentioning that all the geometric phases are zero, which comes from the fact that $(\partial_\auxt \Bra{\phi_m(\auxt)}) \Ket{\phi_m(\auxt)}$ is always an imaginary number, while the coefficients of the eigenstates are all real, which implies that all such terms are zero.
Applying the transformation, we get:
\begin{subequations}
\begin{equation}\label{eq:HTot_IP}
\overline{H}(t) - \bitilde{G}_2(t) = \bitilde{H}_S^\crt\!(t) + \bitilde{H}_{AB}(t) \,,
\end{equation}
with 
\begin{eqnarray}
\bitilde{H}_S^\crt\!(t) &=& \sum_{m=1}^2 \Omega_m(t) \Ket{\tritilde{g}_m(t)}\Bra{\tritilde{e}(t)} \ee^{- \ii 2 \nu' t}  + h.c. \,, \,\,\,\,\,\,\,\,\,\,
\end{eqnarray}
\begin{eqnarray}
\nonumber
\bitilde{H}_{AB}(t) &=& \sum_{m=1}^2 (\Ket{\tritilde{g}_m(t)}\Bra{\tritilde{e}(t)} \ee^{- \ii \nu' t}  + \Ket{\tritilde{e}(t)}\Bra{\tritilde{g}_m(t)} \ee^{\ii \nu' t}  ) 
\otimes \sum_k \eta_k^{(m)}  (\sigma_-^{(k)} \ee^{- \ii \omega t} + \sigma_+^{(k)} \ee^{ \ii \omega t})  \,, \,\,\, \\
\end{eqnarray}
\end{subequations}
with $\Ket{\tritilde{g}_m(t)} = U_2^\dag(t, t_0) \Ket{g_m}$ and $\Ket{\tritilde{e}(t)} = U_2^\dag(t, t_0) \Ket{e}$.

In the new picture, the generator of the time evolution is the following:
\begin{eqnarray}
\hamfintld =  \bitilde{H}_S^\crt (t) + \bitilde{H}_{AB} (t) \,,
\end{eqnarray}
and the relevant approximated dynamics can be evaluated, to the second order, by truncation of the iterated formal solution:
\begin{eqnarray}\label{eq:DefT2}
T(t,t_0) \approx   \opidentity - \ii \int_{t_0}^t \hamfintld(\auxt) \mathrm{d}\auxt - \int_{t_0}^t \mathrm{d}\auxt \int_{t_0}^\auxt \mathrm{d}\auxtt\,  \hamfintld(\auxt) \hamfintld(\auxtt)   \equiv T_2(t,t_0) \,,
\end{eqnarray}
which is essentially the truncation of the Dyson series (expressed without the chronological ordering operator) to the second order.

Moreover, since we have moved to this new picture by removing the adiabatic evolution operator responsible for a perfect population transfer from $\Ket{g_1}$ to $\Ket{g_2}$, remaining in the state $\Ket{g_1}$ is equivalent to undergoing a perfect transition from $\Ket{g_1}$ to $\Ket{g_2}$ in the Schr\"odinger picture.
Therefore, in the new picture the efficiency of the population transfer process through STIRAP pulses is given by the survival probability of the initial state of the three-state system:
\begin{eqnarray}\label{eq:BasicEfficiency}
P(t) = \mathrm{tr} [ T(t,t_0)  \Ket{\psi(0)} \Bra{\psi(0)} \rho_B(0) T^\dag(t) \Ket{\psi(0)} \Bra{\psi(0)} \otimes \opidentity_B ] \,, 
\end{eqnarray}
where $\rho_B(0)$ is the density operator describing the initial configuration of the bath, $\opidentity_B$ is the identity operator of the bath and $T(t,t_0)$ is possibly replaced by $T_2(t,t_0)$. In our case $\Ket{\psi(0)}=\Ket{g_1}$.
In order to better understand \eqref{eq:BasicEfficiency}, consider that the complete time evolution of the initial state $\Ket{g_1} \Bra{g_1} \otimes \rho_B(0)$ in the Schr\"odinger picture is given by:
$\rho_{AB}(t) = {\cal U}_1(t,t_0) \, {\cal U}_2(t,t_0) \, T(t,t_0) \, \Ket{g_1} \Bra{g_1} \otimes \rho_B(0) \, T^\dag(t,t_0) \, {\cal U}_2^\dag(t,t_0) {\cal U}_1^\dag(t,t_0)$. Now, since the target state is $\Ket{g_2}$, irrespectively of the state of the bath, we need to evaluate 
$P(t) = \mathrm{tr}[\rho_{AB}(t)$  $\Ket{g_2}\Bra{g_2} \otimes \opidentity_B]$, which, after performing two cyclic permutations inside the trace functional leads to:
$\mathrm{tr}[ T(t,t_0) \, \Ket{g_1} \Bra{g_1} \rho_B(0) \, T^\dag(t,t_0) \, {\cal U}_2^\dag(t,t_0) {\cal U}_1^\dag(t,t_0)  \Ket{g_2}\Bra{g_2} \otimes \opidentity_B {\cal U}_2(t,t_0)$  ${\cal U}_1(t,t_0)]$,
which is equivalent to \eqref{eq:BasicEfficiency} once it is considered that ${\cal U}_2(t,t_0) \Ket{g_1} = \Ket{g_2}$ and conversely ${\cal U}_2^\dag(t,t_0) \Ket{g_2} = \Ket{g_1}$ so that ${\cal U}_2^\dag(t,t_0) {\cal U}_1^\dag(t,t_0)  \KetBra{g_2}{g_2} \otimes \opidentity_B {\cal U}_1(t,t_0) {\cal U}_2(t,t_0) =$  $\KetBra{g_1}{g_1} \otimes \opidentity_B$.

\section{Results}\label{sec:results}

We now focus on the zero-temperature bath, which means assuming that the spin bath is initially in its ground state $\Ket{\alldown} = \otimes_k \Ket{\downarrow}_k$. Therefore, the complete initial state is $\Ket{g_1}\Ket{\alldown}$.
Since we are considering the approximation $T(t,t_0)\approx T_2(t,t_0)$ according to \eqref{eq:DefT2}, the only transitions considered in our calculations are those involving zero, one or two spin flips in the bath. This implies the following form for the efficiency:
\begin{eqnarray}\label{eq:ExpandedEfficiency}
\nonumber
P(t) &\approx&  |\Bra{\alldown}\Bra{g_1} T_2(t,t_0) \Ket{g_1} \Ket{\alldown}|^2 \\
\nonumber
&+& \Sigma_l | \Bra{\alldownbutone{l}} \Bra{g_1} T_2(t,t_0) \Ket{g_1} \Ket{\alldown}|^2 \\
&+& \Sigma_{j \not= l} |\Bra{\alldownbuttwo{j}{l}} \Bra{g_1} T_2(t,t_0) \Ket{g_1} \Ket{\alldown}|^2 \,, \qquad
\end{eqnarray}
where $\Ket{\alldownbutone{l}}$ is the bath state with all spins in the $\Ket{\downarrow}$ state, except for the $l$-th spin, which is the $\Ket{\uparrow}$ state, while $\Ket{\alldownbuttwo{j}{l}}$ has only the $l$-th and $j$-th spins in the $\Ket{\uparrow}$ state, all the others being in the state $\Ket{\downarrow}$.
The overlaps involving only one spin flip turn out to be zero (see Appendix \ref{app:SecondOrder} for details). The overlaps involving two spin flips involve only second-order terms and, once their squared modulus is evaluated, such terms give rise to fourth-order contributions.
To make the calculation consistent with the truncation of the Dyson series to the second order, only terms up to the second order are to be kept in the probability, which gives the following expression: 
\begin{eqnarray}\label{eq:EfficiencyProportional}
P(t) \approx  1 -  2  \,\, \Re\left[ {\cal J}_{\left\{\eta_k^{(m)}\right\}}(t) \right] \,,
\end{eqnarray}
with
\begin{eqnarray}
\nonumber
{\cal J}_{\left\{\eta_k^{(m)}\right\}}(t) \!\!\! &=& \!\!\! \int_{t_0}^{t} \!\! \mathrm{d} \auxt \int_{t_0}^{\auxt} \!\! \mathrm{d} \auxtt   \bigg[ \ee^{- \ii \int_{\auxt}^{\auxtt} E_+(s)\mathrm{d}s } \, \cos\varphi(\auxt)\cos\varphi(\auxtt) \\
\nonumber
&+& \ee^{- \ii \int_{\auxt}^{\auxtt} E_-(s)\mathrm{d}s } \, \sin\varphi(\auxt)\sin\varphi(\auxtt) \bigg]
\ee^{\ii (\nu' + \omega) (\auxtt-\auxt)} \\ 
&\times&   \sum_k \,  
\left[ \eta_k^{(1)} \cos\theta(\auxt) - \eta_k^{(2)} \sin\theta(\auxt) \right]
\left[ \eta_k^{(1)} \cos\theta(\auxtt) - \eta_k^{(2)} \sin\theta(\auxtt) \right]  \,.
\label{eq:JEtaIntegral}
\end{eqnarray}

We now assume that the quantities $\eta_k^{(m)}$, first defined after \eqref{eq:HabDef} without any constraint, are related in such a way that the ratios $ \eta_k^{(m)} / \eta_1^{(m)}$ do not depend on $m$, so that we can introduce  $\lambda_k \equiv \eta_k^{(m)} / \eta_1^{(m)}$, define the following quantities,
\begin{eqnarray}
\eta \equiv \eta_1^{(1)}\,, \qquad r_\eta \equiv \eta_1^{(2)}/\eta_1^{(1)} \,, \qquad \CorrOne \equiv \sum_{k} \lambda_k^2 \,,
\end{eqnarray}
and recast the probability in the following form:
\begin{eqnarray}\label{eq:SurvProb_J}
P(t) \approx  1 -  2 \eta^2 \CorrOne \,\, \Re[{\cal J}(t)]\,,
\label{eq:FinalProbCut}
\end{eqnarray}
with
\begin{eqnarray}
\nonumber
{\cal J}(t) \!\!\! &=& \!\!\! \int_{t_0}^{t} \!\! \mathrm{d} \auxt \int_{t_0}^{\auxt} \!\! \mathrm{d} \auxtt   \bigg[ \ee^{- \ii \int_{\auxt}^{\auxtt} E_+(s)\mathrm{d}s } \, \cos\varphi(\auxt)\cos\varphi(\auxt) 
+ \ee^{- \ii \int_{\auxt}^{\auxtt} E_-(s)\mathrm{d}s } \, \sin\varphi(\auxt)\sin\varphi(\auxtt) \bigg] \\ 
&\times& \ee^{\ii (\nu' + \omega) (\auxtt-\auxt)}  \,  
\left[ \cos\theta(\auxt) - r_\eta \sin\theta(\auxt) \right]
\left[ \cos\theta(\auxtt) - r_\eta \sin\theta(\auxtt) \right]  \,.
\label{eq:JIntegral}
\end{eqnarray}

On the basis of \eqref{eq:SurvProb_J}, the survival probability of the initial state in the interaction picture, which corresponds to the efficiency of the population transfer in the Schr\"odinger picture, turns out to differ from unity by a term proportional to real part of the the integral ${\cal J}(t)$ (having the dimensions of the square of time), on which we will focus in our further analysis. Every specific correction should take into account the specific value of the square of the quantity $\eta \sqrt{\Lambda}$ (having the dimension of a frequency), which somehow is a cumulative measure of the coupling strength between the three-level system and the whole environment.
It is also interesting to observe that in the RWA the counterpart of ${\cal J}(t)$ would be zero. This can be straightforwardly proven by recalculating the relevant matrix elements. Moreover, it is already well visible from the expression of ${\cal J}(t)$ where all the terms contain rapidly oscillating factors coming from the fact that all the contributions come out as matrix elements of the counter-rotating terms.

\begin{figure}[h!]
\includegraphics[width=0.45\textwidth, angle=0]{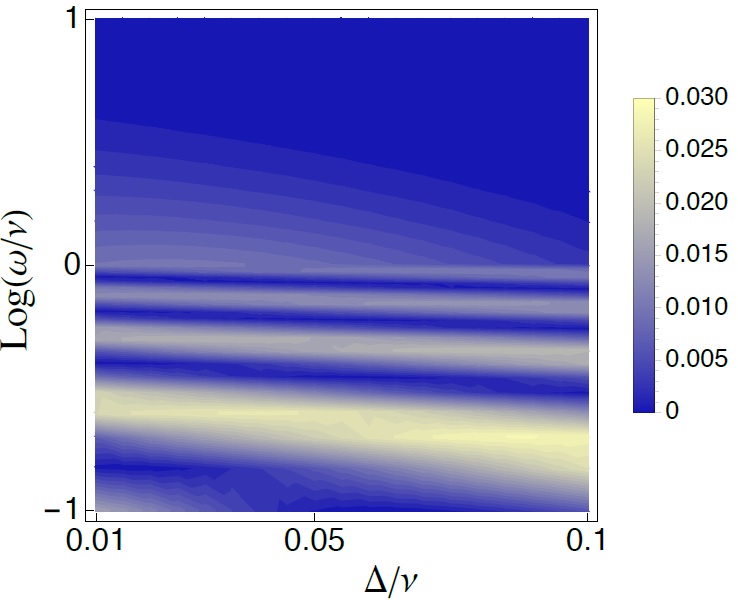} \qquad
\includegraphics[width=0.45\textwidth, angle=0]{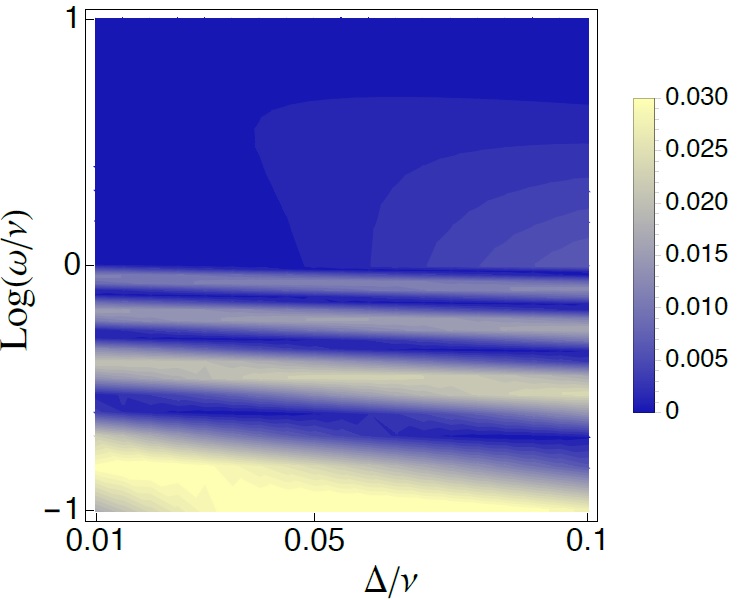} 
\caption{The quantity $2\Re\{{\cal J}(T)\}$ (in units of $\tau^2$) as a function of $\omega$  (in units of $\nu$ and logarithmic scale) and $\Delta$ (in units of $\nu$) in the case $r_\eta=1$ (a) and $r_\eta=-1$ (b). The other parameters are: $\Omega_0\tau=10$, $T/\tau=5$, $\nu\tau=10$.} 
\label{fig:Integral_1}
\end{figure}

\begin{figure}[h!]
\includegraphics[width=0.45\textwidth, angle=0]{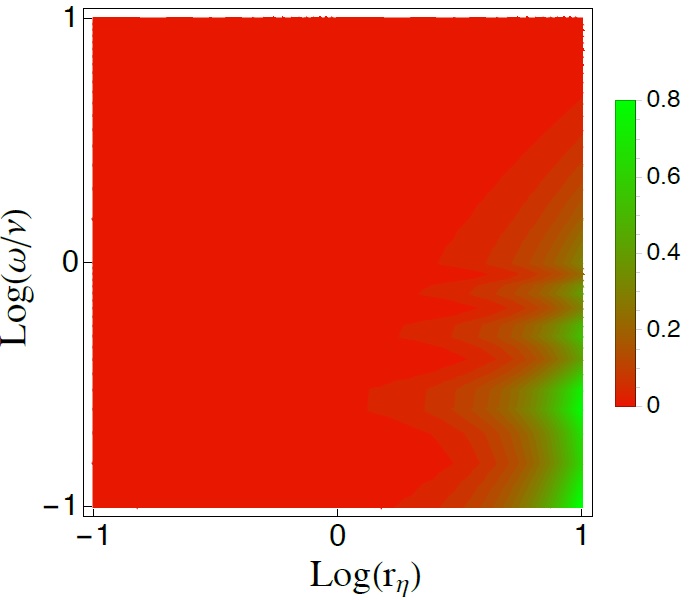} \qquad
\includegraphics[width=0.45\textwidth, angle=0]{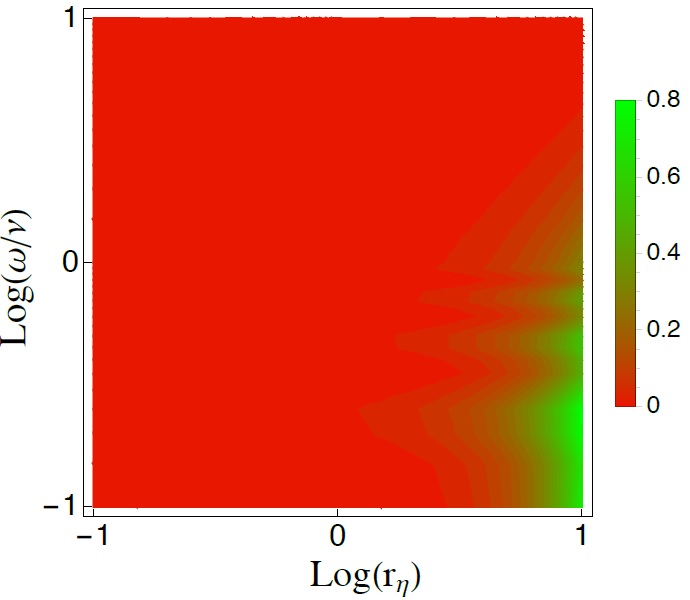}
\caption{The quantity $2\Re\{{\cal J}(T)\}$ (in units of $\tau^2$) as a function of $\omega$ (in units of $\nu$ and logarithmic scale) and $r_\eta$ (in logarithmic scale) for $\Delta=0.01 \, \nu$ (a) and $\Delta=0.05\,\nu (b)$. The other parameters are: $\Omega_0\tau=10$, $T/\tau=5$, $\nu\tau=10$.} \label{fig:Integral_2}
\end{figure}

It is worth mentioning that the assumption $ \eta_k^{(m)} / \eta_1^{(m)}$ independent from the index $m$ is not that restrictive as one could think, since the coupling strengths are supposed to be proportional to a function of the spin distance from the central three-state system, and this proportionality function is supposed to be the same independently from the specific transitions involved, whether $\Ket{e} \leftrightarrow \Ket{g_1}$ or  $\Ket{e} \leftrightarrow \Ket{g_2}$.
Concerning the shape of the pulses, they are usually taken as gaussian, with the peaks occurring at different times. In particular, we assume two pulses centered in $\pm\tau$ and having width $\sim \tau / \sqrt{2}$:
\begin{subequations}
\begin{eqnarray}
\Omega_1 = \Omega_0 \ee^{- (t / \tau + 1 )^2 } \,,    \\
\Omega_1 = \Omega_0 \ee^{- (t / \tau - 1 )^2 }  \,.       
\end{eqnarray}
\end{subequations}
The process is supposed to start at time $t_0 = -T$ (with $T > \tau$) and to finish at $t = T$. In Fig.~\ref{fig:Integral_1} and \ref{fig:Integral_2} is reported the numerically calculated quantity $2\Re\{{\cal J}(T)\}$ as a function of different parameters, which allows to evaluate the efficiency of the population transfer in several regimes: the smaller is the quantity $2\Re\{{\cal J}(T)\}$, the more efficient is the population transfer, according to \eqref{eq:SurvProb_J}. 
In all the plots we have assumed that the parameters of the pulses satisfy the following conditions, which guarantee an optimal transfer in the ideal case: $\Omega_0\tau=10$, $T/\tau=5$, $\nu\tau=10$. 

In Fig.~\ref{fig:Integral_1} it is reported $2\Re\{{\cal J}(T)\}$ as a function of $\omega$ and $\Delta$ for two values of the ratio $r_\eta$, in particular $r_\eta=1$ (a) and $r_\eta=-1$ (b). In both plots it is well visible that the value of $2\Re\{{\cal J}(T)\}$ is always small, never exceeding the value $0.03$, and that for high values of the frequency $\omega$ the corrections become smaller and smaller. For any fixed value of $\Delta$, one can see that varying the value of $\omega$ the quantity $2\Re\{{\cal J}(T)\}$ exhibits an oscillatory behavior, which {\it a posteriori} can be related to the presence of the phase factor $\exp[(\omega+\nu')(\auxtt-\auxt)]$ in the integrand. In fact, in spite of the presence of other functions, the mentioned phase factor is clearly the main rapidly changing factor, whereas trigonometric functions of $\phi$ and $\theta$ are smoothly changing and the phase factor associated to the integral of $E_+$ changes rapidly only in the region between the two peaks. Thus, both the presence of oscillations and the vanishing of ${\cal J}(T)$ for higher values of $\nu$ are traceable back to the oscillatory character of the counter-rotating terms.
By comparing figures (a) and (b), it emerges that whether the value of the ratio parameter $r_\eta$ is $1$ or $-1$ the behavior is pretty similar, though significant differences are present especially in the region corresponding to small values of $\Delta$. 

In Fig.~\ref{fig:Integral_2} the quantity $2\Re\{{\cal J}(T)\}$ is shown as a function of $\omega$ and $r_\eta$ for two values of the detuning $\Delta$, in particular for $\Delta/\nu=0.01$ (a) and $\Delta/\nu=0.05$ (b). It is well visible that the values of the quantity in a particular parameter region are much higher than in the previous plots, reaching the value $0.8$. These high values can be justified by the fact that increasing the value of $r_\eta$ implies increasing the amplitude of the coupling term between the three-state system and the environment. In particular, a high value of $r_\eta$ means dealing with a high value of $\eta_1^{(2)}$ (and consequently all $\eta_k^{(2)}$), which in turn implies a higher value of ${\cal J}(T)$.

In order to fix the ideas about the meaning of our results, let us consider the case where $\eta\sqrt{\Lambda}\tau = 1$, which means $\eta\sqrt{\Lambda}/\Omega_0 \ll 1$, thus keeping valid the perturbation treatment. In such a case, in the greener region of Fig.~\ref{fig:Integral_2} the correction would be about $0.8$, implying an very low efficiency of about $0.2$. On the contrary, in the worst region of Fig.~\ref{fig:Integral_1}, the yellow parts corresponding to $0.03$, the efficiency would be of about $0.97$.

\section{Discussion}\label{sec:discussion}

In this paper we have considered the effects of the interaction with a spin bath on the efficiency of a population transfer realized through a STIRAP process. Our present work is an evolution of the study reported in Ref.~\cite{ref:MiliNapo2023}, where the system-bath interaction has been considered in the RWA. Here, in order to improve the analysis we have also considered the effects of the counter-rotating terms. Nevertheless, since the model analyzed is not exactly solvable and a numerical treatment of it would be challenging, we have faced the problem through a perturbation approach. In particular, second order corrections have been calculated by evaluating the Dyson series truncated to the second order contributions. The complete deviation of the efficiency of the population transfer from unity, according to \eqref{eq:SurvProb_J}, turn out to be proportional to the square of the quantity $\eta\sqrt{\Lambda}$, which somehow plays the role of a perturbation parameter. It is the case to note that $\eta\sqrt{\Lambda} = [\sum_k (\eta_k^{(1)})^2]^{1/2} $ and $r_\eta\eta\sqrt{\Lambda} = [\sum_k (\eta_k^{(2)})^2]^{1/2}$, which correspond somehow to effective strengths of the couplings with the environment involving $\Ket{g_1}\leftrightarrow\Ket{e}$ transitions and  $\Ket{g_2}\leftrightarrow\Ket{e}$ transitions, respectively. Beyond this fact, according to \eqref{eq:SurvProb_J}, the correction to the efficiency of population transfer is proportional to the real part of the integral ${\cal J}(T)$ defined in \eqref{eq:JIntegral}.

Remarkable, in a second order perturbation treatment under the RWA the correction to the efficiency would be zero, thus predicting a perfect population transfer. On the contrary, beyond such a approximation, deviations come up. In fact, our analysis shows that in the second perturbation treatment the counter-rotating terms in the interaction are the only ones really contributing to the corrections. Indeed, in the integrand of ${\cal J}(T)$ appear only contributions proportional to the phase factor $\exp[(\omega+\nu')(\auxtt-\auxt)]$ traceable back the rapidly oscillating terms in the interaction (namely, the counter-rotating terms), while no term proportional to a phase factor associated to a lower frequency $\omega-\nu'$ (the rotating terms) is present. This clearly implies that in a model involving RWA, the correction is zero, which is perfectly in agreement with the results of Ref.~\cite{ref:MiliNapo2023} where the efficiency always exhibits a {\it plateau} in the weak coupling limit and up to the weak-intermediate coupling regime.

The numerical evaluation of the integral ${\cal J}(T)$ reported in Fig.~\ref{fig:Integral_1} and \ref{fig:Integral_2} shows that in the range of the parameters analyzed the effects of the environment are mainly negligible, except for specific regions where the ratio $r_\eta$ assumes high values, since this implies that the coupling constants $\eta_k^{(2)}$ are significant. 
Further numerical results (not reported in this manuscript because the would essentially provide uniform tablets corresponding to zero values) show in a clear way that higher and higher values of $\nu$ imply smaller and smaller corrections, due to the fact that counter-rotating terms become more and more rapidly oscillating and thus ineffective. Summing up, negligibility of the effects of the environment is obtained for high values of $\omega$ (according to the plates shown in the manuscript) or high values of $\nu$ (according to other simulations), provided $r_\eta$ is not too large. Obviously the values of $\eta\sqrt{\Lambda}$ are to be kept small in order to maintain the validity of our analysis based on perturbation theory.

It is worth concluding with two more comments concerning the comparison with the results in Ref.~\cite{ref:MiliNapo2023}. First, since our second-order perturbation treatment is valid only in the weak coupling regime, we cannot obtain corrections in the strong coupling regime where a generalized quantum Zeno effect were predicted to occur. Second,  the shape of the pulses here is properly gaussian, while in Ref.~\cite{ref:MiliNapo2023} different expressions have been considered. Nevertheless, in spite of the more complicated analytical expression, those pulses are essentially gaussian too, in the sense that they only slightly differ form the gaussian counterparts, which makes the two situations reasonably comparable.

\section*{Acknowledgments}
The authors gratefully acknowledge financial support from the FFR2021 grant from the University of Palermo, Italy.

\appendix

\section{Adiabatic Approximation}\label{app:Adiabatic}

Here we recall the proof of the adiabatic theorem~\cite{ref:Messiah,ref:Griffiths}. Consider a system subjected to time-dependent Hamiltonian $H(t)$ which can be diagonalized at every instant of time in such a way that $H(t) \Ket{\phi_{nj}(t)} = E_n(t) \Ket{\phi_{nj}(t)}$, where each energy eigenspace has a possible degeneracy taken into account through the index $j$. Expanding the state of the system in terms of such instantaneous eigenstates, 
\begin{eqnarray}
\Ket{\psi(t)} = \sum_{nj} a_{nj}(t) \ee^{-\ii \int_{t_0}^t E_n(s) \mathrm{d}s} \Ket{\phi_{nj}(t)}\,,
\end{eqnarray}
and inserting this expression in the Schr\"odinger equation leads to a set of equations for the coefficients $a_{nj}$.
Now, from $\PBraKet{\phi_{nj}(t)}{\phi_{n'j'}(t)}=\delta_{nn'}\delta_{jj'}$ and $\PBra{\phi_{nj}(t)} H(t) \PKet{\phi_{n'j'}(t)} =E_n(t) \delta_{nn'}\delta_{jj'}$ we get: $\partial_t (\PBraKet{\phi_{nj}(t)}{\phi_{n'j'}(t)}) = 0$ and  $\partial_t(\PBra{\phi_{nj}(t)} H(t) \PKet{\phi_{n'j'}(t)}) = 0$ unless $n=n'$ and $j=j'$. After some algebra, one gets that for $n\not=k$ the following relation occurs: $\PBraKet{\phi_{nj}(t)}{\dot{\phi}_{kl}(t)} = \PBra{\phi_{nl}(t)} \dot{H}(t)\PKet{\phi_{kj}(t)} / (E_k(t)-E_n(t))$.
On this basis once can eventually write down the following set of equations:
\begin{eqnarray}
\nonumber
\dot{a}_{nj}(t) &=& - \PBraKet{\phi_{nj}(t)}{\dot{\phi}_{nj}(t)} \, a_{nj}(t) \, - \sum_{l\not=j} \, \PBraKet{\phi_{nj}(t)}{\dot{\phi}_{nl}(t)} \, a_{nl}(t) \\
&-& \sum_{k\not=n}\sum_l \, \ee^{-\ii \int_0^t (E_k(s) - E_n(s)) \mathrm{d} s} \, \frac{\PBra{\phi_{nl}(t)} \dot{H}(t)\PKet{\phi_{kj}(t)} }{E_k(t)-E_n(t)}   \, a_{kl}(t)  \,.
\end{eqnarray}
Under the assumption of very small matrix elements of the operator $\dot{H}$ one can neglect the terms in the second line, and for non degenerate eigenspaces one eventually gets $\dot{a}_{nj} \approx - \PBraKet{\phi_{nj}}{\dot{\phi}_{nj}} \, a_{nj}$ or, equivalently, after using $\PBraKet{\chi}{\dot\chi}=-\PBraKet{\dot\chi}{\chi}$,
\begin{eqnarray}
\dot{a}_{nj}(t) \approx \PBraKet{\dot{\phi}_{nj}(t)}{\phi_{nj}(t)} \, a_{nj}(t) \,.
\end{eqnarray}

The complete evolution is then given by:
\begin{eqnarray}
\Ket{\psi(t)} \approx \sum_{nj} a_{nj}(0) \,\, \ee^{\int_{t_0}^t \PBraKet{\dot{\phi}_{nj}(s)}{\phi_{nj}(s)} \mathrm{d}s} \,\, \ee^{-\ii \int_{t_0}^t E_n(s) \mathrm{d}s} \, \Ket{\phi_{nj}(t)}\,,
\end{eqnarray}
where the quantity $\PBraKet{\dot{\phi}_{nj}(s)}{\phi_{nj}(s)}$, known as the geometric phase, is an imaginary number. Therefore, if the coefficients of the expansion of $\Ket{\phi_{nj}(s)}$ with respect to a given basis are all real, the phase turns out to be the sum of real numbers, which is then supposed to be equal to zero. This is the case for the eigenstates of the STIRAP Hamiltonian considered above.

\section{Second order term for the zero-temperature bath}\label{app:SecondOrder}

In the calculation of the corrections up to the second order we will need some quantities. Since $\PBraKet{g_1}{\tritilde{\phi}(t)}  = \Bra{g_1}U_2^\dag(t, t_0)\Ket{\phi}=\Bra{\phi}U_2(t, t_0)\Ket{g_1}^*$, considered the expressions of the states of the adiabatic basis and the fact that at the initial time ($t=t_0$) of the counterintuitive sequence $\Ket{0}=\Ket{g_1}$, $\Ket{+}=\Ket{e}$ and $\Ket{-}=\Ket{g_2}$, one easily finds:
\begin{eqnarray}
\BraKet{g_1}{\tritilde{e}(t)} &=& 0 \,, \\
\label{eq:g1g1}
\BraKet{g_1}{\tritilde{g}_1(t)} &=& \cos\theta(t) = \Omega_2(t) / \Omega_L(t) \\ 
\label{eq:g1g2}
\BraKet{g_1}{\tritilde{g}_2(t)} &=& -\sin\theta(t) = -\Omega_1(t) / \Omega_L(t)  \\ 
\label{eq:ee}
\BraKet{\tritilde{e}(t)}{\tritilde{e}(t')}  &=& \ee^{- \ii \int_{t}^{t'} E_+(s)\mathrm{d}s } \, \cos\varphi(t)\cos\varphi(t') 
+ \ee^{- \ii \int_{t}^{t'} E_-(s)\mathrm{d}s } \, \sin\varphi(t)\sin\varphi(t') \,, \qquad
\end{eqnarray}
where $\theta(t)$ and $\varphi(t)$ given by \eqref{eq:thetadef} and  \eqref{eq:phidef}.  

Since $\BraKet{g_1}{\tritilde{e}(t)}=0$, all the first order terms turn out to be zero, when it is calculated any probability to find the three-state system in the state $\Ket{g_1}$. Indeed, whatever the bath states $\Ket{\Psi_B}$ and $\Ket{\Psi_B'}$, the matrix elements $\Bra{\Psi_B}\Bra{g_1} \bitilde{H}_S^\crt (t) \Ket{g_1}\Ket{\Psi_B}$ and $\Bra{\Psi_B}\Bra{g_1} \bitilde{H}_{AB}(t) \Ket{g_1}\Ket{\Psi_B}$ give rise to terms proportional to the overlap $\BraKet{g_1}{\tritilde{e}(t)}$ (or its adjoint), which is zero.

Let us now then consider the second order contributions.
First of all, observe that the matrix element $\PBra{\alldown}\Bra{g_1} \hamfintld(\auxt) \hamfintld(\auxtt) \Ket{g_1} \PKet{\alldown}$ admits potential contributions coming only from the terms $\bitilde{H}_S^\crt (\auxt)\bitilde{H}_S^\crt (\auxtt)$ and $\bitilde{H}_{AB}(\auxt) \bitilde{H}_{AB}(\auxtt)$. Indeed, $\bitilde{H}_S^\crt (\auxt)\bitilde{H}_{AB} (\auxtt)$ and  $\bitilde{H}_{AB} (\auxt) \bitilde{H}_S^\crt (\auxtt)$ involve one spin flip coming from $\bitilde{H}_{AB}$ while we are considering the matrix element between $\Ket{g_1} \Ket{\alldown}$ and itself. That said, let us then focus on the other two terms:
\begin{eqnarray}
\nonumber
&&\Bra{\alldown} \Bra{g_1} \bitilde{H}_S^\crt (\auxt)\bitilde{H}_S^\crt (\auxtt) \Ket{g_1} \Ket{\alldown} = 
\ee^{2\ii\nu' (\auxt-\auxtt)} \BraKet{e(\auxt)}{e(\auxtt)} \, \\ 
&&\qquad \times \,
\sum_{mm'} \Omega_m(\auxt)\Omega_{m'}(\auxtt) \BraKet{g_1}{\tritilde{g}_m(\auxt)} \BraKet{\tritilde{g}_{m'}(\auxtt)}{g_1}   = 0\,, 
\end{eqnarray}
whose being zero is based on \eqref{eq:g1g1} and \eqref{eq:g1g2}, and
\begin{eqnarray}
\nonumber
&&\Bra{\alldown} \Bra{g_1} \bitilde{H}_{AB}(\auxt) \bitilde{H}_{AB}(\auxtt) \Ket{g_1} \Ket{\alldown} = 
\ee^{\ii (\nu' + \omega) (\auxtt-\auxt)} \BraKet{e(\auxt)}{e(\auxtt)}  \,  \\ 
&&\qquad \times \BraKet{g_1}{\tritilde{g}_m(\auxt)} \BraKet{\tritilde{g}_{m'}(\auxtt)}{g_1} 
\sum_k \eta_k^{(m)} \eta_k^{(m')} \,,
\end{eqnarray}
which, after using \eqref{eq:g1g1}, \eqref{eq:g1g2} and \eqref{eq:ee}, gives the integrand in \eqref{eq:JEtaIntegral}.


\begin{thebibliography}{999}







\bibitem{ref:Messiah} A. Messiah, Quantum Mechanics (Dover, Mineola, 1995).

\bibitem{ref:Griffiths} D. J. Griffiths, Introduction to Quantum Mechanics (Cambridge University Press, 2016).


\bibitem{ref:Lidar2018} Albash T. and Lidar D. A. Adiabatic quantum computation. {\em Rev. Mod. Phys.} {\bf 2018}, {\em 90}, 015002. 

\bibitem{ref:Santos2020} Santos A. C. and Sarandy M. S.  Sufficient conditions for adiabaticity in open quantum systems. {\em Phys. Rev. A 102}, {\bf 2020}, {\em 052215}. 

\bibitem{ref:Ilin2021} Il'in, N.,  Aristova A., and Lychkovskiy O.  Adiabatic theorem for closed quantum systems initialized at finite temperature. {\em Phys. Rev. A }{\bf 2021}, {\em 104}, L030202. 

\bibitem{ref:Pys2022} Pyshkin P. V. , Da-Wei Luo, and Lian-Ao Wu Self-protected adiabatic quantum computation.  {\em Phys. Rev. A} {\bf 2022}, {\em 106}, 012420.  

\bibitem{ref:Coello2022} Coello P\'erez E. A., Bonitati J., Lee D., Quaglioni S., and Wendt K. A., Quantum state preparation by adiabatic evolution with custom gates. {\em Phys. Rev. A} {\bf 2022}, {\em 105}, 032403. 


\bibitem{ref:STIRAP_rev1} Vitanov N. V., Fleischhauer M., Shore B. W. and Bergmann K., Coherent manipulation of atoms molecules by sequential laser pulses. {\em Adv. At. Mol. Opt. Phys.} {\bf 2001},  {\em 46}, 55.

\bibitem{ref:STIRAP_rev2} Vitanov N. V., Halfmann T., Shore B. W. and Bergmann K. {\em Ann. Rev. Phys. Chem.} {\bf 2001}, {\em 52}, 763.

\bibitem{ref:STIRAP_rev3} Bergmann K., Theuer H. and Shore B. W. {\em Rev. Mod. Phys.} {\bf 1998}, {\em 70}, 1003.

\bibitem{ref:STIRAP_rev4} Kr\'al P., Thanopulos I., and Shapiro M. {\em Rev. Mod. Phys.} {\bf 2007}, {\em 79}, 53.

\bibitem{ref:STIRAP_rev5} Bergmann K. {\it et al.} {\em J. Phys. B: At. Mol. Opt. Phys.} {\bf 2019}, {\em 52}, 202001.

\bibitem{ref:Blekos2020} Blekos K., Stefanatos D., Paspalakis E. Performance of superadiabatic stimulated Raman adiabatic passage in the presence of dissipation and Ornstein-Uhlenbeck dephasing. {\it Phys. Rev. A} {\bf 2020} {\it 102}, 023715.

\bibitem{ref:Ahmadinouri2020} Ahmadinouri F., Hosseini M., Sarreshtedari F. Stimulated Raman adiabatic passage: Effects of system parameters on population transfer. {\it Chem. Phys.} {\bf 2020}, {\it 539}, 110960.

\bibitem{ref:Cinins2022} Cinins A., Bruvelis M., Bezuglov N. N. {\em J. Phys. B: At. Mol. and Opt. Phys.}  {\bf 2022}, {\em 55(23)}, 234003. 

\bibitem{ref:Dogra2022} Dogra S. and Paraoanu G. S. Perfect stimulated Raman adiabatic passage with imperfect finite-time pulses. {\it J. Phys. B: At. Mol. Opt. Phys.} {\bf 2022},  55, 174001.

\bibitem{ref:Liu2023} Liu K., Sugny D., Chen X., Gu\'erin S. {\em Entropy} {\bf 2023},  {\em 25(5)}, 790. 

\bibitem{ref:Genov2023} Genov G. T., Rochester S., Auzinsh M., Jelezko F., Budker D. {\em J. Phys. B: At. Mol. and Opt. Phys.} {\bf 2023}, {\em 56(5)},054001. 

\bibitem{ref:Ni2008} Ni K.-K., Ospelkaus S., de Miranda M. H. G., Pe\^{a}er A.,  Neyenhuis B., Zirbel J. J., Kotochigova S., Julienne P. S., Jin D. S., Ye J. {\em Science} {\bf 2008}, {\em 322} , 231. 

\bibitem{ref:Danzl2008} Danzl J. G., Haller E., Gustavsson M., Mark M. J., Hart R., Bouloufa N.,  Dulieu O., Ritsch H., N\"agerl H.-C. {\em Science} {\bf 2008}, {\em 321}, 1062. 

\bibitem{ref:Danzl2010} Danzl J. G., Mark M. J., Haller E., Gustavsson M., Hart R., Aldegunde J.  Hutson J. M. and N\"agerl H.-C., {\em Nat. Phys.} {\bf 2010}, {\em 6}, 265. 

\bibitem{ref:Halfmann2007} Klein J., Beil F., and Halfmann T. {\em Phys. Rev. Lett.} {\bf 2007}, 99, 113003.  

\bibitem{ref:Alexander2008} Alexander A. L., Lauro R.,  Louchet A., Chaneliere T. and  Le Gouet J. L. {\em Phys. Rev. B} {\bf 2008},  {\em 78}, 144407.

\bibitem{ref:Golter2014}  Golter D. A. and Wang H. {\em Phys. Rev. Lett.} {\bf 2014},  {\em 112}, 116403. 

\bibitem{ref:Yale2016} Yale C. G., Heremans F. J., Zhou B. B., Auer A., Burkard G. and Awschalom D. D. {\em Nature Photonics} {\bf 2016}, {\em 10}, 184.

\bibitem{ref:Wolfowicz2021} Wolfowicz, G., Heremans, F.J., Anderson, et al., Quantum guidelines for solid-state spin defects, {\em Nat Rev Mater} {\bf 2021}, {\em 6}, 906925.

\bibitem{ref:Zhou2017}  Zhou B.,  Baksic A., Ribeiro H. et al. {\em Nature Phys.} {\bf 2017}, {\em 13}, 330. 

\bibitem{ref:Baksic2016} Baksic A., Ribeiro H. and Clerk A. A. {\em Phys. Rev. Lett.} {\bf 2016}, {\em 116}, 230503.

\bibitem{ref:Varguet2016} Varguet H., Rousseaux B., Dzsotjan D., Jauslin H. R.,  Gu\'erin S., Colas des Francs G. {\em Opt. Lett.} {\bf 2016} {\em 41}, 4480.

\bibitem{ref:Castellini2018} Castellini A., Jauslin H. R., Rousseaux B., Dzsotjan D., Colas des Francs G., Messina A. and Gu\'erin S. {\em Eur. Phys. J. D} {\bf 2018}, {\bf 72}, 223.

\bibitem{ref:Kubo2016} Kubo Y. {\em Nat. Phys.} {\bf 2016}, {\em 12} 212. 

\bibitem{ref:Xu2016} Xu H. K., Song C., Liu W. Y., Xue G. M., Su F. F., Deng H., Ye Tian, Zheng D. N., Siyuan Han, Zhong Y. P.,  Wang H., Yu-xi Liu and Zhao S. P.  {\em Nat. Commun.}{\bf 2016} {\em 7}, 11018.  

\bibitem{ref:Kumar2016} Kumar K. S., Veps\"al\"ainen A., Danilin S. and Paraoanu G. S. {\em Nat. Commun.} {\bf 2016}, {\em 7}, 10628. 

\bibitem{ref:Soresen2006} Soresen J. L., Moller D., Iversen T., Thomsen J. B., Jensen F., Staanum P., Voigt D. and  Drewsen M. {\em New. J. Phys.} {\bf 2006}, {\em 8}, 261. 

\bibitem{ref:Higgins2017} Higgins G., Pokorny F., Zhang C.,  Bodart Q. and Hennrich M., {\em Phys. Rev. Lett.}  {\bf 2017} {\em 119}, 220501. 

\bibitem{ref:Fedoseev2021} Fedoseev V., Luna F., Hedgepeth I., L\"offler W. and Bouwmeester D. Stimulated Raman Adiabatic Passage in Optomechanics. {\it Phys. Rev. Lett.} {\bf 2021} {\it 126}, 113601.

\bibitem{ref:Guery2019}  Gu\'{e}ry-Odelin D., Ruschhaupt A., Kiely A., Torrontegui E., Mart\'{i}nez-Garaot S., and Muga J. G. {\em Rev. of Mod. Phys.} {\bf 2019}, {\em 91}, 045001.

\bibitem{ref:Vitanov2020} Vitanov N. V. High-fidelity multistate stimulated Raman adiabatic passage assisted by shortcut fields. {\em Phys. Rev. A} {\bf 2020},  {\em 102}, 023515.

\bibitem{ref:Evange2023} Evangelakos V., Paspalakis E., Stefanatos D. {\em Phys. Rev. A} {\bf 2023} {\em 107(5)} ,052606.  

\bibitem{ref:Stefanatos2022} Stefanatos D., Paspalakis E. {\em Philosophical Transactions of the Royal Society A: Mathematical, Physical and Engineering Sciences} {\bf 2022}, {\em 380(2239)}, 20210283.  

\bibitem{ref:Messikh2022} Messikh C., Messikh A. {\em EPL} {\bf 2022}, {\em 140(4)}, 48003. 

\bibitem{ref:Stefanatos2020} Stefanatos D., Blekos K., Paspalakis E., {\em Applied Sciences (Switzerland)} {\bf 2020}, {\em 10(5)},1580. 

\bibitem{ref:Genov2013} Genov G. T. and Vitanov N. V. {\em Phys. Rev. Lett.} {\bf 2013},  {\em 110}, 133002.

\bibitem{ref:Yatsenko2014} Yatsenko L. P., Shore B. W. and Bergmann K., {\em Phys. Rev. A} {\bf 2014}, {\em 89}, 013831. 

\bibitem{ref:Vitanov1997} Vitanov N. V. and Stenholm S. {\em Phys. Rev. A} {\bf 1997}, {\em 56}, 1463. 

\bibitem{ref:Petruccione} Breuer H.-P. and Petruccione F., {\it The Theory of Open Quantum Systems\/} (Oxford University Press, Oxford, 2002).

\bibitem{ref:Gardiner} Gardiner C. W. and Zoller P., {\it Quantum Noise\/} (Springer-Verlag, Berlin, 2000).

\bibitem{ref:Ivaniv2005} Ivanov P. A., Vitanov N. V. and Bergmann K. {\em Phys. Rev. A} {\bf 2005}, {\em, 72}, 053412. 

\bibitem{ref:New1} Akram M. J. and Saif F., {\em J. Russ. Laser Res.} {\bf 2014}, {\em 35}, 547. 

\bibitem{ref:New2} Sukharev M. and Malinovskaya S. A.,   {\em Phys. Rev. A} {\bf 2018}, {\em 86}, 043406. 

\bibitem{ref:Mathisen2018} Mathisen T. and Larson J., {\em Entropy} {\bf 2018}, {\em 20}, 20. 

\bibitem{ref:DaviesSpohn}  Davies E. B. and Spohn H.  {\it J. Stat. Phys.} {\bf 1978}, {\it 19}, 511.

\bibitem{ref:Scala2010}  Scala M.,  Militello B.,  Messina A. and Vitanov N. V. {\em Phys. Rev. A} {\bf 2010}, {\em 81}, 053847. 

\bibitem{ref:Scala2011} Scala M.,  Militello B.,  Messina A. and Vitanov N. V. {\em Phys. Rev. A} {\bf 2011}, {\em 83}, 012101. 


\bibitem{ref:Onizhuk2021} Onizhuk M., Miao K. C., Blanton J. P., He Ma, Anderson C. P., Bourassa A., Awschalom D. D.  and Galli G., {\em Phys. Rev. X Quantum} {\bf 2021}, {\em 2}, 010311. %

\bibitem{ref:MiliNapo2023} Militello B. and Napoli A. Adiabatic Manipulation of a System Interacting with a Spin Bath, {\em Entropy} {\bf 2023}, {\em 15}, 2028.

\bibitem{ref:Fischer2007} Fischer J. and  Breuer H.-P. {\em Phys. Rev. A} {\bf 2007}, {\em 76}, 052119.

\bibitem{ref:Ferraro2008} Ferraro E., Breuer H.-P.,  Napoli A., Jivulescu M. A. and Messina A. {\em Phys. Rev. B} {\bf 2008}, {\em 78}, 064309.

\bibitem{ref:Bhatta2017} Bhattacharya S., Misra A., Mukhopadhyay C. and Pati A. K.  {\em Phys. Rev. A} {\bf 2017}, {\em 95}, 012122.











\end{thebibliography}
\end{document}